\documentclass[useAMS,usenatbib]{mn2e}
\usepackage{natbib}
\usepackage{color,graphicx} 
\usepackage{amsmath}        
\usepackage{amsfonts}       
\usepackage{amssymb}        
\usepackage{wasysym}        

\begin{document}
\author[K.~Liu et al.]
{K.~Liu,$^{1,2}$\thanks{kliu@mpifr-bonn.mpg.de} C.~G.~Bassa,$^{3}$
G.~H.~Janssen,$^{3}$ R.~Karuppusamy,$^{1}$ J.~McKee,$^{4}$
M.~Kramer,$^{1,4}$
\newauthor K.~J.~Lee,$^{5}$ D.~Perrodin,$^{6}$ M.~Purver,$^{4}$ S.~Sanidas,$^{7,3}$ R.~Smits,$^{3}$ B.~W.~Stappers,$^{4}$
\newauthor P.~Weltevrede$^{4}$ and W.~W.~Zhu,$^{1}$\\
  $^{1}$Max-Planck-Institut f\"{u}r Radioastronomie, Auf dem H\"{u}gel
  69, D-53121 Bonn, Germany \\
  $^{2}$Station de radioastronomie de Nan\c{c}ay, Observatoire de
  Paris, CNRS/INSU, F-18330 Nan\c{c}ay, France \\
  $^{3}$ASTRON, the Netherlands Institute for Radio Astronomy, Postbus 2, NL-7990 AA, Dwingeloo, The
  Netherlands \\
  $^{4}$University of Manchester, Jodrell Bank Centre for Astrophysics,
  Alan Turing Building, Manchester M13 9PL, UK\\
  $^{5}$KIAA, Peking University, Beijing 100871, P.R. China \\
  $^{6}$INAF - Osservatorio Astronomico di Cagliari, Via della Scienza
5, I-09047 Selargius (CA), Italy\\
  $^{7}$Anton Pannekoek Institute for Astronomy, University of Amsterdam,
Science Park 904, NL-1098 XH Amsterdam, The Netherlands \\
  }

\title[Single pulses of PSR~J1713+0747]{Variability, polarimetry, and timing properties of single pulses from
PSR~J1713+0747 using the Large European Array for Pulsars}

\maketitle

\begin{abstract}
Single pulses preserve information about the pulsar radio emission
and propagation in the pulsar magnetosphere, and understanding the
behaviour of their variability is essential for estimating the
fundamental limit on the achievable pulsar timing precision. Here we
report the findings of our analysis of single pulses from
PSR~J1713+0747 with data collected by the Large European Array for
Pulsars (LEAP). We present statistical studies of the pulse properties that
include distributions of their energy, phase and width. Two modes of
systematic sub-pulse drifting have been detected, with a periodicity
of 7 and 3 pulse periods. The two modes appear at
different ranges of pulse longitude but overlap under the main peak
of the integrated profile. No evidence for pulse micro-structure is
seen with a time resolution down to 140\,ns. In addition, we show
that the fractional polarisation of single pulses increases with
their pulse peak flux density. By mapping the probability density of
linear polarisation position angle with pulse longitude, we reveal
the existence of two orthogonal polarisation modes. Finally, we find
that the resulting phase jitter of integrated profiles caused by
single pulse variability can be described by a Gaussian probability
distribution only when at least 100 pulses are used for integration.
Pulses of different flux densities and widths contribute
approximately equally to the phase jitter, and no improvement on
timing precision is achieved by using a sub-set of pulses with a
specific range of flux density or width.
\end{abstract}


\section{Introduction} \label{sec:intro}
Millisecond pulsars (MSPs) that were spun up in accreting binary
systems to reach rotational periods $\lesssim30$\,ms \citep{acrs82},
are noted for their highly precise timing behaviour
\citep[e.g.][]{abb+15,dcl+16,rhc+16}. Their short and stable
rotational period make them excellent tools for probing tiny
spacetime perturbations and performing gravity experiments,
including tests of General Relativity with great precision
\citep[e.g.][]{ksm+06,wnt10}, stringent constraints on alternative
theories of gravity \citep[e.g.][]{fwe+12,afw+13}, probes of neutron
star equations-of-state \citep[e.g.][]{dpr+10,opr+10}, and the
ongoing search for gravitational waves in the nanohertz regime
\citep[e.g.][]{ltm+15,srl+15,abb+16}.

The success of the aforementioned timing experiments is attributed
to both the regular rotation of the pulsars, and their stable
integrated pulse profiles formed by averaging over tens of thousands
of periods. Nevertheless, it has been known since the discovery of
pulsars that the pulsed emission from every single rotation (thus
single pulses) of a pulsar is highly variable. This was first
noticed in canonical pulsars and more recently in MSPs
\citep[e.g.][]{jak+98,sc12,lkl+15}. A consequence of such
variability is the so-called pulse phase jitter phenomenon in
integrated profiles, which for a given integration time places a
fundamental limit on the achievable timing precision on short
timescales. Timing precision has already reached the jitter-limited
regime for a few MSPs \citep{lkl+11,sc12}, and this limit is
expected to be reached for many more when the next generation of
radio telescopes (e.g., the Square Kilometre Array) comes online
\citep{lvk+11,jhm+15}. Most recent timing analysis has started to take
into account the effect of phase jitter when modelling the noise in the timing data \citep[e.g.][]{ls15,zsd+15,cll+16}.
Detailed studies of single pulse variability in MSPs are crucial for building a comprehensive
understanding of this phenomenon, if we are to push beyond this
limitation. Such investigations will also provide input for the
efforts to either model or mitigate jitter noise in timing data
\citep{ovh+11,ick+15}.

The origin of pulsar radio emission is associated with plasma
processes in the highly-magnetised pulsar magnetosphere
\citep[e.g.][]{cr77,cor79,cjd04}, the understanding of which has
still been elusive. Single pulse data preserve information of the
intensity, polarisation, and even waveform (if dual-polarization
Nyquist sampled time series are recorded) of the emission from every
single rotation. Studying single pulses can shed light on the nature
of the pulsar emission mechanism, by, e.g., revealing the
fundamental units of coherent radiation \citep{cor76a,gil85,jap01},
distinguishing different modes of polarised emission
\citep{gl95,es04}, characterising the temporal variability in pulse
intensity \citep{rs75,gmg03,wes06}, and so forth. Studying pulsars that display the
drifting sub-pulse phenomenon can also assist in determining the
viewing geometry and provide insight into the structure of the
emission region \citep{dc68,rs75,ran86,qlz+04}. The vast majority of
single pulse emission studies has been mostly of canonical (normal,
non-recycled) pulsars \citep[see][for an overview]{lg06}. Extending
this work to include MSPs will establish a bridge between the
understanding of emission physics in canonical pulsars and that in
MSPs, so as to see if a common theory or model of pulsar radio
emission can be applied.

A small fraction of MSPs emit occasional giant radio pulses, which
have been studied in detail
\citep{cstt96,kbm+06b,kni07,zps+13,bpd+15}. However, investigations
into ordinary single pulses of MSPs have been carried out only for a
limited number of bright sources
\citep{jak+98,es03a,sc12,bil12,ovb+14,sod+14,lkl+15}, one of which
is PSR~J1713+0747. This pulsar is one of the most precisely timed
pulsars and has been included in current pulsar timing array
campaigns to detect gravitational waves in the nanohertz frequency
range \citep{vlh+16}. The binary system it inhabits is also an ideal
test laboratory for alternative theories of gravity \citep{zsd+15}.
\cite{es03a} showed that there is clear modulation in the pulses
from PSR~J1713+0747 and that it varies across the pulse profile and
as a function of observing frequency. In the fluctuation spectra,
they showed that the pulsar exhibited two broad maxima corresponding
to fluctuations at 0.17 and 0.35\,cycles-per-period (cpp).
However, the longitude dependence and a correlation
with drifting sub-pulses could not be established due to the lack of
sensitivity. \cite{sc12} have shown that single pulses from
PSR~J1713+0747 are highly variable in phase, which already limits
its timing precision with the current observing sensitivity. Thus,
understanding single pulse variability of this pulsar in order to
potentially mitigate its contribution to the timing noise is clearly
necessary and will be the focus of this work.

The rest of this paper is structured as follows. In
Section~\ref{sec:obs}, we provide the details of the observation and
preprocessing of the data. Section~\ref{sec:res} presents the results
from our data analysis on single pulse variability, polarisation
and timing properties. We conclude in Section~\ref{sec:conclu} with a
brief discussion and prospects for future work.

\section{Observations} \label{sec:obs}
Investigations of single pulses from MSPs have been greatly
restricted by their comparatively low flux density in the radio
band, as well as the lack of availability of single pulse data. This
is because of the large data volumes and precision timing typically
being performed on integrated pulse data. The Large European Array
for Pulsars (LEAP) performs monthly simultaneous observations at
1.4\,GHz of twenty-three MSPs with the five 100-m class radio
telescopes in Europe \citep{bjk+16}. Coherently combining the
voltage data from the radio telescopes delivers the sensitivity
equivalent to a single dish with diameter up to 195\,m and gain up to
5.7\,K\,Jy$^{-1}$. For all pulsars observed, the product of
coherently-added voltages can be used to generate single pulse data
with full polarisation information and variable time resolution. In
addition to LEAP's high sensitivity, the monthly observing campaign
increases the chances of achieving a high signal-to-noise (S/N)
detection, especially for low-DM pulsars, whose flux density varies
dramatically due to diffractive interstellar scintillation.
Therefore, LEAP provides a database which is ideal for single pulse
studies of MSPs.

For the study in this paper, we selected 15\,minutes of archived
data obtained on MJD~56193 (23 September 2012) when a bright scintle was
caught in the LEAP frequency band. The observation was conducted simultaneously with
the Effelsberg Radio Telescope, the Nan\c{c}ay Radio telescope, and
the Westerbork Synthesis Radio Telescope (WSRT)\footnote{The Lovell Telescope at Jodrell Bank
was under maintenance during the observation.}. At Effelsberg, the
PSRIX pulsar backend was used in baseband mode to record
8$\times$16\,MHz bands of two orthogonal polarisations at the
Nyquist rate and with 8-bit sampling \citep{lkg+16}. The eight bands are centred at 1340, 1356, 1372, 1388,
1404, 1420, 1436, and 1452\,MHz. At Nan\c{c}ay, we used
the NUPPI instrument \citep{dbc+11} to record baseband data from four
of the eight bands (1388, 1404, 1420, 1436\,MHz) with an identical
setup. At the WSRT, the PuMa-II system \citep{ksv08} was used to
record 8-bit baseband data of 8$\times$20\,MHz bands, with central
frequencies at 1342, 1358, 1374, 1390, 1406, 1422, 1438 and 1454\,MHz\footnote{The WSRT bands overlap by 4\,MHz, so the coherent addition
later used 16\,MHz out of the 20\,MHz.}. At both Effelsberg and
the WSRT, the data were later copied to disks and shipped to the Jodrell
Bank Observatory (JBO), where they were installed into the LEAP central storage
cluster. The data from Nan\c{c}ay were directly transferred to JBO via
internet. A full description of the LEAP observational setup can be found
in \cite{bjk+16}.

Preprocessing of the data was then carried out on the CPU cluster at
JBO using a software correlator developed specifically for LEAP
(Smits et al. in prep.). Each observation was divided into 3-min
sub-integrations for correlation purposes. To calibrate the
polarisation of the voltage data from each individual telescope, we
first dedispersed and folded the data to form an integrated profile
from the entire observation. Using the polarisation profile of
PSR~J1713+0747 \citep[from][]{stc99} obtained from the European
Pulsar Network (EPN) database\footnote{http://www.epta.eu.org/epndb/}, we
applied a template matching technique to the observed profile in
order to measure the receiver properties, assuming the full
reception model as in \cite{van04a}. The corresponding Jones
matrices were then calculated and used to calibrate the voltage
data. In order to correct for the difference in voltage phase
response between different backends, we used the observation of the
phase calibrator J1719+0817 (performed immediately before the pulsar
observation) to measure the phase offsets across the entire band,
and then applied them to the data of PSR~J1713+0747. As it was
a very bright observation of the pulsar, we used the pulsar data
itself to measure the fringes (relative time and phase offsets)
between different sites. With the fringe solutions, the software
correlator produced coherently added voltage data with an effective
bandwidth of 112\,MHz centred at 1388\,MHz\footnote{Data from the
1452\,MHz sub-band were not included due to a receiver cutoff at
Effelsberg.}, which were then processed with the DSPSR software
\citep[for details, see][]{vb11} to extract the single pulse data.
The data were coherently dedispersed using a dispersion measure (DM)
of $15.9891\,\rm cm^{-3}pc$ (which is the default value used at JBO
to de-disperse PSR~J1713+0747 data). Single pulse data were written
to disks with a maximum time resolution of 140\,ns (corresponding to
32,768 phase bins within a period). More details of our data
processing pipeline are described in \cite{bjk+16}.

\begin{figure}
\centering
\includegraphics[scale=0.47,angle=-90]{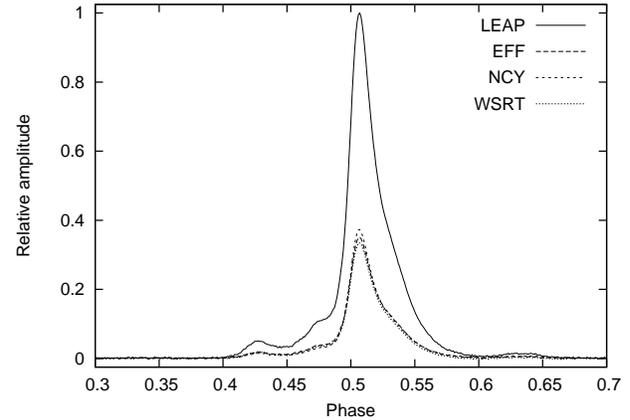}
\caption{Total intensity profile of PSR~J1713+0747 on MJD~56193 from
LEAP data integrated over the entire 15-min observation, compared
with those from single telescope data. The amplitude of the profiles from individual
telescopes are relative to that of the LEAP profile. The peak signal-to-noise ratios are
896, 314, 333, 300 for LEAP, Effelsberg (EFF), Nan\c{c}ay (NCY), and Westerbork (WSRT),
respectively, which corresponds to an addition with a 95\%
coherency. The peak of the LEAP profile locates at phase 0.507. \label{fig:prof}}
\end{figure}

\begin{figure}
\centering
\includegraphics[scale=0.35,angle=-90]{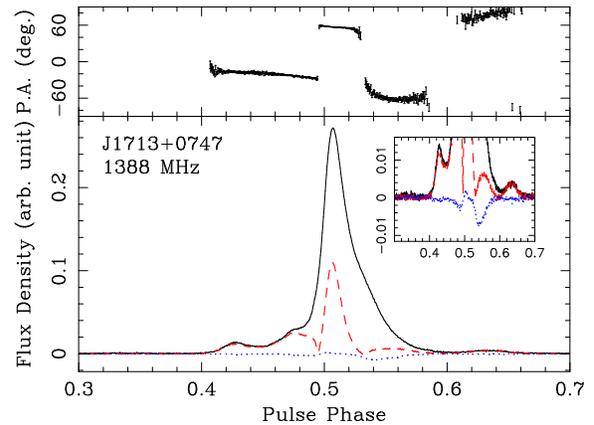}
\caption{Polarisation profile and linear polarisation position angle (P.A.) of PSR~J1713+0747. 
In the lower panel, the solid (black), dashed (red) and dotted (blue) lines 
represent the total intensity ($I$), linear ($L$) and circular ($V$) 
components, respectively. The inner panel shows an enlarged view of the polarisation profile.
The time resolution used is 2.3\,$\mu$s.\label{fig:polprof}}
\end{figure}

Fig.~\ref{fig:prof} shows the total intensity profile from the LEAP
data compared with those from each of the single telescopes. A
calculation based on signal-to-noise ratio measurements shows a
coherency of 95\% for the LEAP addition. The polarisation profile as
well as the swing of linear polarisation position angle (P.A.) from
the LEAP data shown in Fig.~\ref{fig:polprof} is consistent with the
results from previous work \citep{ovhb04,ymv+11}. We also used the receiver
properties measured from PSR~J1713+0747 to calibrate PSR~J1600$-$3053 data
obtained on the same epoch, and had consistent result as shown in the EPN
database.

The variability of single pulse emission is commonly investigated
using a set of statistical tools. The modulation index is used to
indicate the level of intensity variation. The fluctuation (power)
spectra of pulse intensity as a function of pulse number and
rotational phase (pulse stack), can identify periodicities in the
emission. The longitude-resolved fluctuation spectrum (LRFS)
resolves the fluctuations as a function of rotational phase. The
two-dimensional fluctuation spectrum (2DFS) is calculated by
performing a two-dimensional Fourier Transform of the pulse stack,
which is widely used to study the systematic drifting of sub-pulses.
Details of these statistical tools can be found in e.g.,
\cite{es03a} and \cite{wes06}. The software tools used to conduct single pulse
analysis in this paper are part of the PSRSALSA project \citep{wel16}, and are freely
available online\footnote{https://github.com/weltevrede/psrsalsa}.

\section{Results} \label{sec:res}
\begin{figure}
\centering
\includegraphics[scale=0.35,angle=-90]{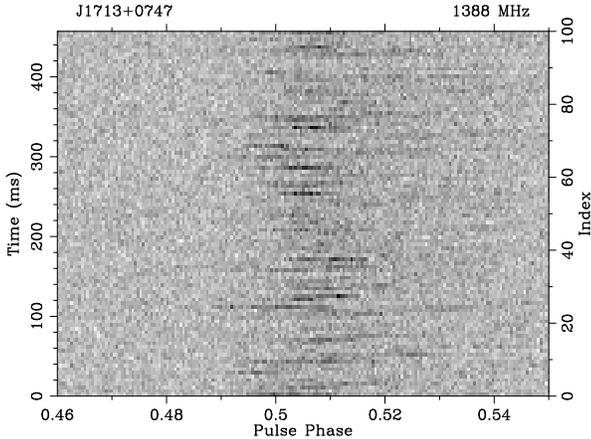}
\caption{An example of a pulse stack containing one hundred pulses.
The time resolution used here is 2.3\,$\mu$s. The figure illustrates the pulse-to-pulse
variation which is analysed quantitatively further below (see Fig.~\ref{fig:lrfs+2dfs}).
\label{fig:samp100}}
\end{figure}

The observation recorded a total of 196,915 single pulses,
approximately 75\% of which were detected with a peak
signal-to-noise ratio (S/N) higher than 5 based on a time resolution
of 9\,$\mu$s. The highest peak S/N is 36, a factor of three greater
than the maximum recorded in \cite{sc12}. Fig.~\ref{fig:samp100}
shows an example stack of one hundred consecutive pulses, which
shows clear intensity variation among the pulses. During the 15-min
observation, the flux density of the pulsar dropped by approximately
10\% due to interstellar scintillation.

\subsection{Single pulse properties} \label{ssec:sglvar}
\begin{figure}
\centering
\includegraphics[scale=0.65,angle=-90]{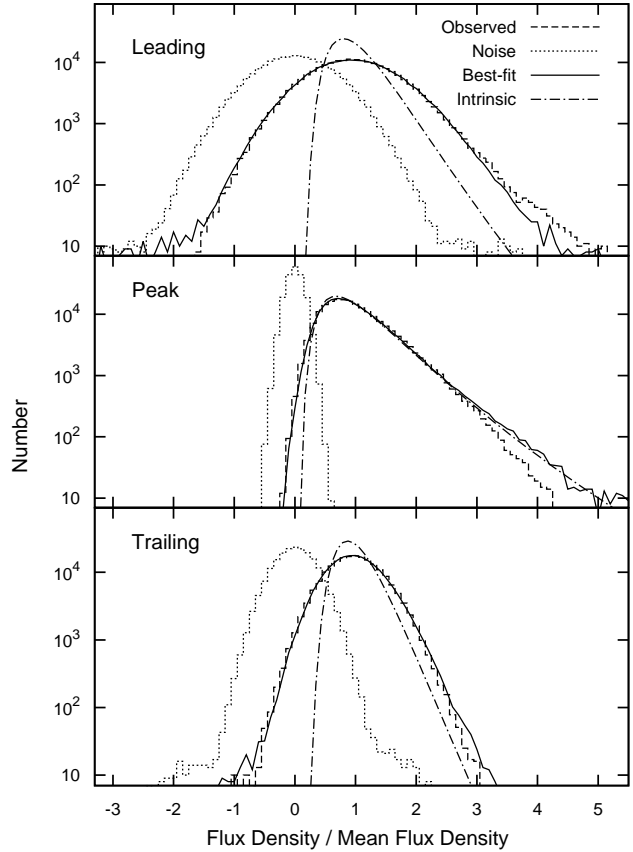}
\caption{Pulse energy distributions in phase ranges (detailed in
Table~\ref{tab:powspec}) corresponding to the leading edge, the
peak, and the trailing edge of the integrated profile.
The dashed, dotted, solid, and dashed-dotted lines represent the
observed pulse energy distribution, the noise energy distribution,
the best-fit to the observed distribution, and the modelled
intrinsic pulse energy distribution with a log-normal function,
respectively. \label{fig:specs}}
\end{figure}

The energy distribution of single pulses from PSR~J1713+0747
has been studied in previous work \citep{sc12,sod+14}, based on
either the peak flux density or the total flux density from a chosen
on-pulse region. Here we used the single pulse data to study the
energy distribution in three phase ranges that correspond to the
leading edge, the peak, and the trailing edge of the integrated
profile (see Fig.~\ref{fig:specs}). The observed pulse
energy was calculated by summing the intensities within the
corresponding phase range defined in Table~\ref{tab:powspec} after
subtracting the off-pulse mean. The corresponding noise energy
distributions were obtained from an equally long phase range in the
off-pulse region. To compensate for the variation in flux density
caused by scintillation, the data were first re-scaled to ensure
that the average pulse energy in subsequent 30-s intervals remains
constant. The observed energy distribution was modelled by
convolving an intrinsic distribution, here assumed to be log-normal,
with the observed noise distribution via a procedure detailed in
\cite{wws+06} and \cite{wel16}. The log-normal probability density
function is defined as
\begin{equation}
\rho(x)=\frac{1}{\sqrt{2\pi}x\sigma}{\rm exp}\left[-\frac{({\rm
ln}x-\mu)^2}{2\sigma^2}\right].\label{eq:log-normal}
\end{equation}
For each region, the observed and the modelled intrinsic pulse
energy distributions are shown in Fig.~\ref{fig:specs}. It can be
seen that the three regions have a different intrinsic pulse energy
distribution, that of the peak region being broadest. A
Kolmogolov-Smirnov (K-S) test (summarised in Table 1) shows that in
all three cases the intrinsic pulse energy can be well described by
an intrinsically log-normal distribution.

\begin{table}
\centering \caption{Phase (see Fig.~\ref{fig:prof} to correspond to the
integrated profile) ranges defined for the leading edge, the peak, and the
trailing edge of the integrated profile, $p$-values from a K-S test on the modelled and
the observed pulse energy distributions in these three regions, and
the best-fit modelling parameters as defined in Eq.~\ref{eq:log-normal}.
None of these $p$-values indicates a significant difference between
the modelled and observed distributions. \label{tab:powspec}}
\begin{tabular}[c]{ccccc}
\hline
&Phase range &$p$-value &$\mu$ &$\sigma$\\
\hline
Leading &$(0.400, 0.500)$ &0.99 &-0.090 &0.37\\
Peak &$(0.500, 0.515)$    &0.60 &-0.12 &0.52\\
Trailing &$(0.515, 0.615)$ &0.67 &-0.052 &0.30\\
\hline
\end{tabular}
\end{table}

\begin{figure}
\centering
\includegraphics[scale=0.35,angle=-90]{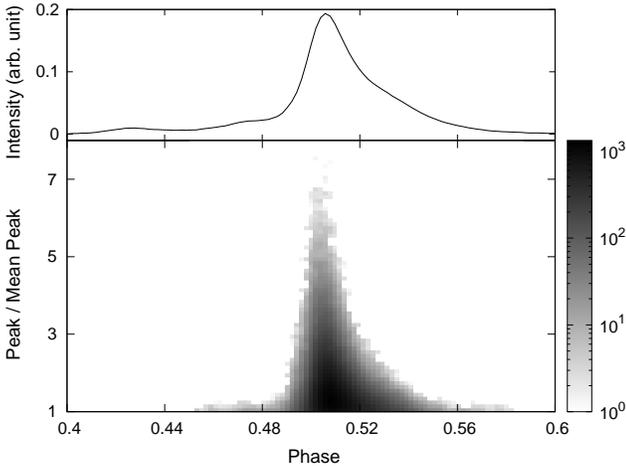}
\caption{Number density distribution (lower panel) of pulses with
  respect to their relative peak flux densities and phase of pulse peak,
  compared with the integrated profile (upper panel). Here we used a time
  resolution of 9\,$\mu$s for the data.
\label{fig:phavf}}
\end{figure}

In Fig.~\ref{fig:phavf}, we plot the number density distribution of
pulses with respect to their relative peak flux densities and phases
of the pulse peak. We note that pulses with higher peak flux
densities tend to be located within a narrow phase range that
includes the maximum of the integrated profile. Lower-amplitude
pulses can occur in a wider phase range which extends more to the
trailing edge of the integrated profile. This confirms the finding
in \cite{sc12}, and indicates that the phase distribution of these
pulses is not symmetric. More investigations into the impact on
timing follow in Section~\ref{ssec:sgltim}. The occurrence rate of
bright pulses whose relative peak flux densities are larger than 3
was found to be constant during the observation.

\begin{figure}
\centering
\includegraphics[scale=0.35,angle=-90]{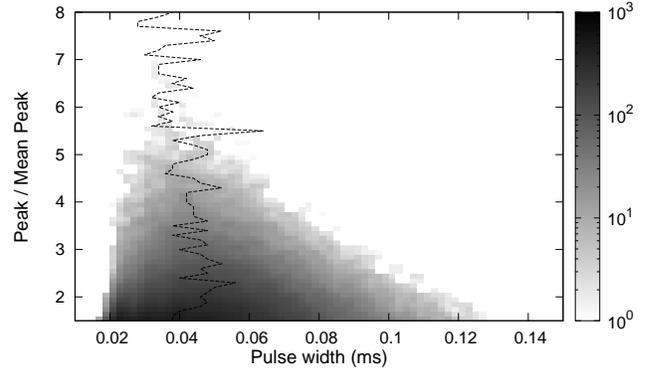}
\caption{Number density distribution of pulses with respect to their
relative peak flux density and width. The peak bins were selected
based on a time resolution of 9\,$\mu$s. The dashed line shows the
pulse width corresponding to the maximum number density for a given
peak flux density. \label{fig:w50vspk}}
\end{figure}

It was shown in \cite{lkl+15} that the bright single pulses from the
MSP~J1022+1001 have a preferred pulse width for any given peak flux
density. To investigate whether a similar feature is present in the
pulses of PSR~J1713+0747, in Fig.~\ref{fig:w50vspk} we plotted the
number density distribution of pulses with respect to their relative
peak flux density and width. Here we define pulse width as the full
width at 50\% of the peak amplitude (i.e., $W_{\rm 50}$), as in
e.g., \cite{lhk+10}. We found that a pulse width of roughly 0.04\,ms
is favoured for the bright single pulses, which is smaller than the
$W_{\rm 50}$ of the integrated profile (0.11\,ms).

\begin{figure}
\centering \includegraphics[scale=1.0,angle=-90]{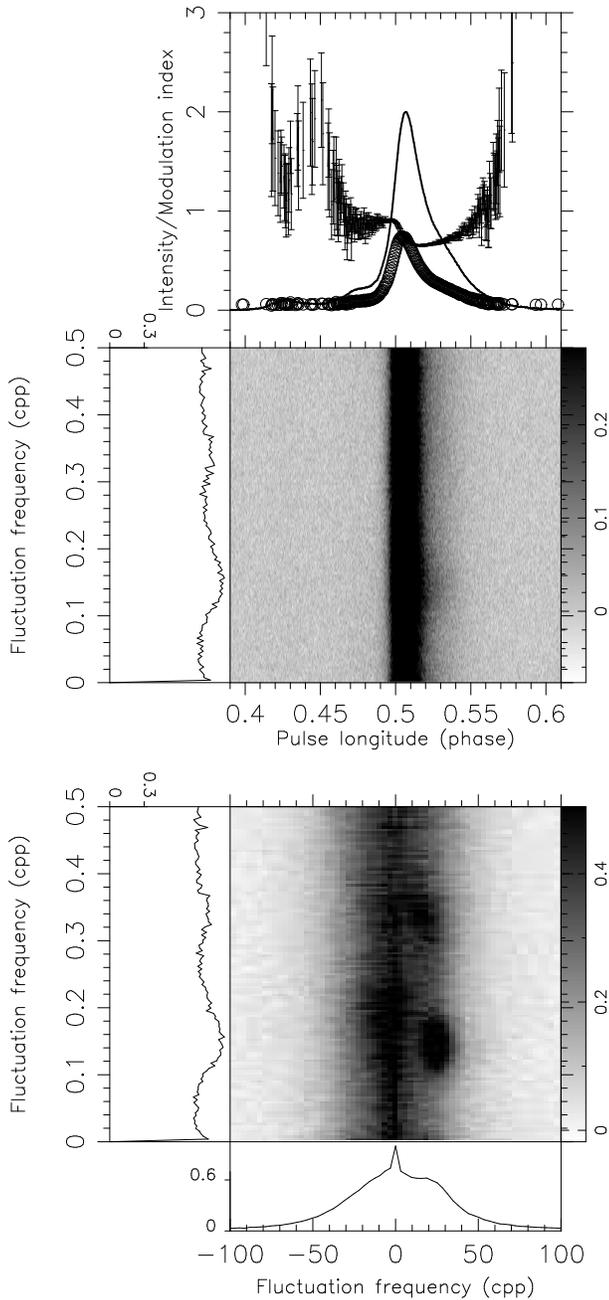}
\caption{Statistical results from analysing the pulse stack from the entire 15-min
  observation. The top panel shows the integrated
  profile (solid line), longitude-resolved modulation index (points
  with error bars), and longitude-resolved standard deviation (open
  circles). The LRFS is plotted below this panel. Below the LRFS, the 2DFS is shown
  and the power in the 2DFS is vertically integrated and then normalised with respect
  to the peak value to produce the bottom plot. Both
  the LRFS and 2DFS are horizontally integrated and then normalised with respect
  to the peak value to produce the left side-panels of the spectra. The right side-panels
  label the colour scale (in arbitrary unit) in the LRFS and 2DFS, respectively. \label{fig:lrfs+2dfs}}
\end{figure}

\begin{figure}
\centering
\includegraphics[scale=0.7,angle=-90]{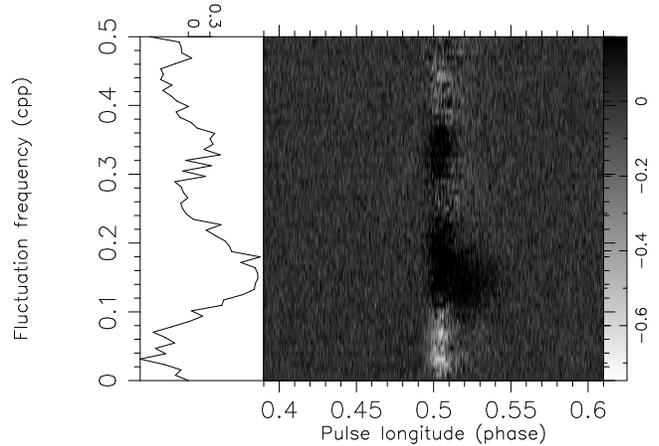}
\caption{The same LRFS as in Fig.~\ref{fig:lrfs+2dfs},
but with the mean power subtracted in each column. The spectrum is horizontally integrated
and then normalised with respect to the peak value, producing the side panel. The right side-panel
label the colour scale in the LRFS with power in arbitrary unit.
\label{fig:lrfs-hk}}
\end{figure}

To investigate the variability of single pulses, we have calculated
the longitude-resolved modulation index, the LRFS, and the 2DFS
based on the entire length of the observation
(Fig.~\ref{fig:lrfs+2dfs}). Here we used the same definitions as in
\cite{wes06}. The modulation index as a function of pulse phase was
found to be consistent with the results in \cite{es03a} and
\cite{sc12}, and yields a significantly improved resolution.
It is maximised at the extreme edges of the integrated
profile, which is commonly seen in canonical pulsars \citep[e.g.][]{wes06}.
The asymmetric distribution of the modulation index also indicates that
the intensity variation is different in the leading and trailing edges of the integrated
profile, as expected from the results in Fig.~\ref{fig:specs}. In
the LRFS shown in Fig.~\ref{fig:lrfs+2dfs}, two maxima are visible
at 0.14 and 0.34\,cpp. This detection confirms the
discovery of periodic intensity modulation by \cite{es03a} with much
higher significance. For better visualisation of the two modes of
periodicity, we subtracted the mean power in each spectrum (i.e.
column) and plotted the modified LRFS in Fig.~\ref{fig:lrfs-hk}. It
can be seen that both periodicities occur at the phase of the main
peak of the integrated profile. The mode with periodicity of
0.14\,cpp extends further to the trailing edge of the integrated
profile, where no significant power was detected corresponding to
the periodicity of 0.34\,cpp. While similar phenomena are known to
occur in canonical pulsars \citep[e.g.][]{kjv02}, this is the first
time a phase-dependent periodicity of intensity variations has been
reported for a MSP.

In the 2DFS, the two local maxima at 0.14 and 0.34\,cpp are clearly
offset from the vertical axis of zero, indicating the association of
the two periodicities with systematic drifting of emission power in
pulse phase. Following the description in \cite{wes06}, we have
measured the horizontal separation of the drifting in pulse
longitude ($P_2$) and their vertical separation in pulse periods
($P_3$) for the two modes, which are ($15^{+2}_{-6}$\,deg,
$6.9\pm0.1$\,$P$) and ($23^{+2}_{-14}$\,deg, $2.9\pm0.1$\,$P$),
respectively. Here $P$ denotes the rotational period of the pulsar.
Both modes show power over a broad range of $P_3$, meaning that the
intensity variation does not follow one constant periodicity. The
discovery of two drifting modes together with their overlap in pulse
phase, suggests that either the two modes appear at the same time at
the same rotational phase, or that the modes alternate. Following
the description in \cite{ssw09}, a potential mode switch
corresponding to the two periodicities was searched for by
calculating the time-resolved 2DFS with a resolution from 10 to 50
pulses. The brightest part of the observation was examined, but our
sensitivity was insufficient to confirm if any transitions occurred
on these short timescales.


\begin{figure}
\centering \includegraphics[scale=0.33,angle=-90]{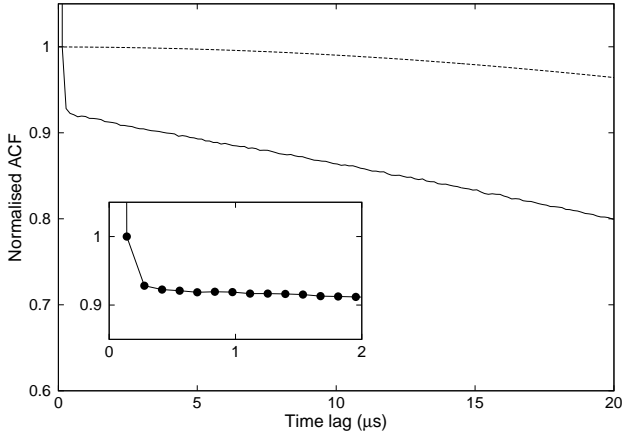}
\caption{Averaged ACF of all single pulses (solid line) compared
with the ACF of the averaged profile (dashed line). The inner panel
shows a zoomed version of the plot in the $0-2$\,$\mu$s range. Here
we retained the highest time resolution of 140\,ns for the profiles.
\label{fig:autocorr}}
\end{figure}

Previous high-time-resolution observations have shown that periodic
micro-structure in pulse emission is common for canonical pulsars
\citep{cor79}. The duration of each micropulse can range from
several hundreds down to several microseconds \citep{ccd68,bs78}.
However, no such phenomenon has thus far been found in MSPs
\citep[e.g.][]{jak+98}. To search for periodic micro-structure in
PSR~J1713+0747, following the definition in \cite{lkwj98}, we
calculated the averaged autocorrelation function (ACF) over all
single pulses compared with the ACF of the averaged profile in
Fig.~\ref{fig:autocorr}. The highest time resolution of 140\,ns in
the data was retained to optimise the sensitivity to narrow
features. Here, because the zero-lag bin of the averaged ACF from
single pulses has a power exceeding 20 times of the other bins, we
used the power in the first non-zero-lag bin to normalise the ACF in
both cases. The comparison between the ACFs suggests that the single
pulses are significantly narrower than the averaged profile. The slope of the averaged
ACF over all single pulses changes at around 280\,ns (the second non-zero-lag bin).
Similar behaviour in the same type of ACF has already been witnessed in canonical pulsars \citep{han72,lkwj98}.
However, here we cannot rule out the cause by instrumental effect as the change of slope
was also seen in the ACF calculated based on the off-pulse region. No
reappearing local maximum has been seen from either ACF, ruling out
the presence of periodic micro-structure. We further examined the
ACFs of the brightest 100 single pulses, based on both peak flux
density and total flux density, and again found no
evidence of micro-structure. In order to rule out potential DM
smearing that may lead to the non-detection, we refolded the
brightest 10 pulses with different DMs ranging from $15.9879$ to
$15.9983\,\rm cm^{-3}pc$\footnote{The central value of the range is
$15.9935\,\rm cm^{-3}pc$, the DM on the observation epoch derived
from \cite{dcl+16}. The chosen range also covers the default DM
value used in dedispersing the data.}, in increments of
$4\times10^{-4}$ which corresponds to 120\,ns smearing time for our
observing frequency and bandwidth. The ACFs again showed no sign of
micro-structure.

\subsection{Single pulse polarisation} \label{ssec:sglpol}
\begin{figure*}
\centering
\includegraphics[scale=0.6,angle=-90,bb=200 0 560 800]{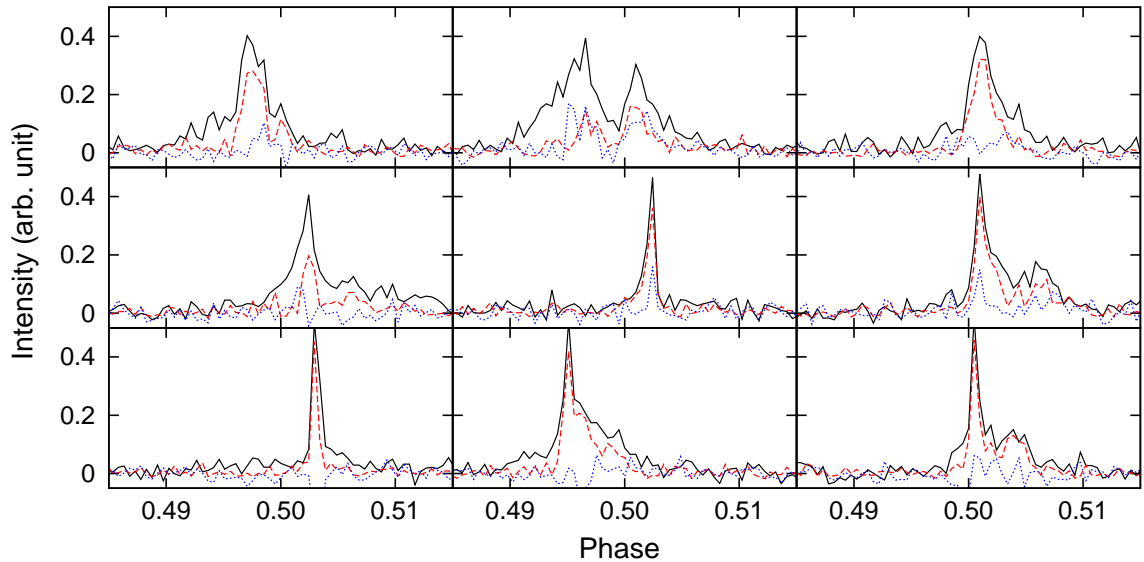}
\caption{Polarisation profiles of the nine highest-S/N pulses. The
definition of the lines is the same as in Fig.~\ref{fig:polprof}.
The time resolution in this plot is 2.3\,$\mu$s.
\label{fig:samples}}
\end{figure*}

\begin{figure}
\centering
\includegraphics[scale=0.35,angle=-90]{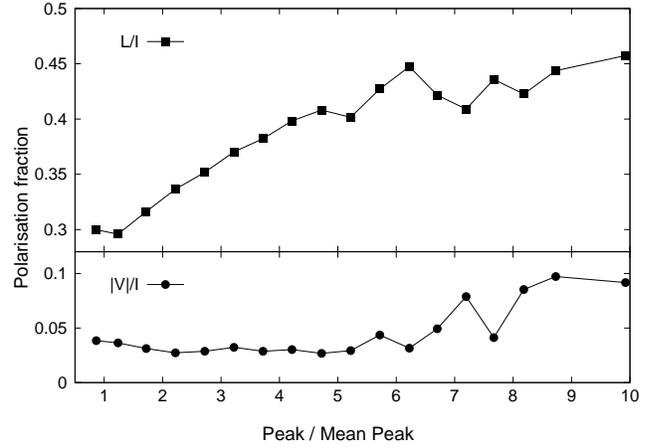}
\caption{Polarisation fractions of linear (top) and circular
(bottom) component, as a function of relative peak flux density of
single pulses. \label{fig:fracpol}}
\end{figure}

The polarisation properties of single pulses from MSPs have not been
widely investigated by previous studies due to both the lack of
sensitivity and hardware constraints. With the coherently added
voltage data as recorded by LEAP, it was possible to retain full
polarisation information for all single pulses with proper
calibration. Example polarisation profiles can be found in
Fig.~\ref{fig:samples}, where we show the nine highest-S/N pulses.
We found that all these pulses show significant linear polarisation,
whose fraction with respect to the total intensity appears to be
higher when compared with the integrated profile
(Fig.~\ref{fig:polprof}). This indicates that the polarisation
fraction may depend on the brightness of the pulses. For further
investigation, in Fig.~\ref{fig:fracpol} we grouped the single
pulses based on their peak flux density and calculated the
polarisation fractions from their averages. The bias in linear and
circular polarisation was corrected following \cite{ss85} and
\cite{tjb+13}. The rise of polarisation fraction for both $L$ and
$V$ with increasing peak flux density is clearly shown. Note that
similar dependence has already been seen in a handful of canonical
pulsars and another MSP \citep{mgm09,ovb+14}, while an opposite
dependence was also noticed in one case \citep{xkjw94}.

\begin{figure}
\centering
\includegraphics[scale=0.4,angle=-90]{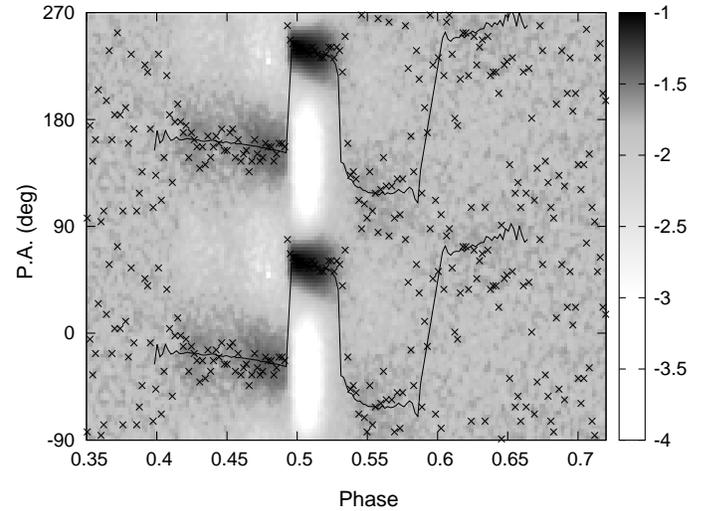}
\caption{Longitude-resolved probability density distribution of P.A.
values from single pulses in log-scale. The values were selected if
the corresponding linear polarisation exceeds 3-$\sigma$. The
crosses represent the maximum of probability for each phase bin. The
solid line shows the P.A. curve from the integrated profile. Note
that between phase 0.59 and 0.60 there is no available P.A
measurement due to the lack of sufficient linear intensity. All P.A.
values were plotted twice (with 360\,deg separation) for clarity.
\label{fig:PActr}}
\end{figure}

The P.A. curve of the integrated profile in Fig.~\ref{fig:polprof}
shows a few discontinuities of approximately 90\,deg difference,
which usually indicates the existence of orthogonal polarisation
modes (OPMs). Following \cite{gl95} and \cite{ovb+14}, in
Fig.~\ref{fig:PActr} we calculated the probability density
distribution of P.A. measured from single pulse data, and compared
with the P.A. swing obtained from the integrated profile
(Fig.~\ref{fig:polprof}). Here we only used samples with a
corresponding linear intensity exceeding 3-$\sigma$ of detection.
For a given phase, we formed the number density distribution with
each count weighted by the inverse of its measurement variance, and
normalised the distribution by its sum. The P.A. values of the single pulses between phase
0.494 and 0.530 (mode 2) were found to be rotated by 90\,deg
compared to adjacent phase ranges (mode 1), which corresponds to the
two jumps in the P.A. curve of the integrated profile. Both the P.A. swing in the integrated
profile and the P.A. distribution of the single pulses show a rapid change around phase 0.6.
However, the low degree of linear polarisation makes it impossible to distinguish between
a smooth transition or a discontinuity. In either case the P.A. changes by less than 90\,deg.


\begin{figure}
\centering
\includegraphics[scale=0.38,angle=-90]{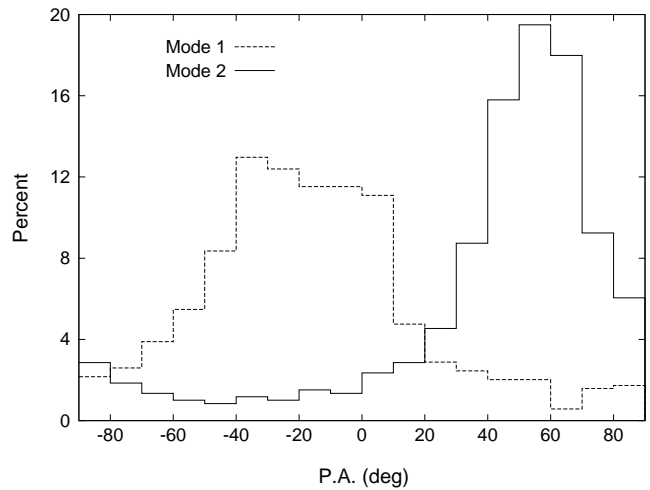}
\caption{Number densities of P.A. (normalised by the sum) from pulse
phase 0.475 (dashed line) and 0.522 (solid line) which correspond to
different orthogonal polarisation modes. The linear intensities in
the integrated profile are the same at the two phases.
\label{fig:PAdist}}
\end{figure}

Within some ranges of pulse phase, in particular those corresponding
to mode 1, the most-likely value of P.A. is widely scattered,
covering a range of nearly 90\,deg. For several selected phase
bins, we checked the number density distributions of P.A. which
turned out to be spread over at least a few tens of degrees, broader
than what is expected purely from instrumental noise (less than ten
degrees). The widths of the distributions grow as the linear
intensity decreases. Such phenomena have already been noticed in
many canonical pulsars \citep[e.g.,][]{scr+84,ms98,kkj+02}. To
examine if the distribution of P.A. are different in the two modes,
we chose pulse phases of the same linear intensity in the integrated
profile, and plotted the distribution of P.A. values with
measurement error less than 7\,deg\footnote{The measurement
uncertainties were obtained by standard error propagation using the
variance of Stokes $Q$ and $U$. It is expected to give the
1-$\sigma$ error, when the S/N of the linear intensity exceeds 3
which corresponds to an error of approximately 9.6\,deg in P.A.
\citep{wk74}.}. The result is shown in Fig.~\ref{fig:PAdist}, where
the distribution from mode 1 is seen to be broader than in mode 2.
This also indicates the existence of an intrinsic P.A. distribution
instead of a single value. A recovery of the intrinsic distribution,
however, would require a detailed modelling of the instrumental
noise which additionally takes into account the covariance of the
noise component in the measured Stokes $Q$ and $U$ \citep{van10}.

\subsection{Impact on timing} \label{ssec:sgltim}
The variability of single pulses introduces a variation in the
integrated profile. This manifests itself as stochastic fluctuations
in the derived pulse time-of-arrivals (TOAs), which is commonly
known as phase jitter. \cite{sc12} have shown that the timing
precision of PSR~J1713+0747 on short timescales can be greatly
limited by phase jitter. The subsequent jitter noise in the timing
data scales as the square-root of number of averaged pulses, as
expected from the theoretical model \citep[e.g.][]{cd85}. Following the method
described in \cite{lkl+11}, we measured jitter noise of 494\,ns in our data
for a 10-s integration time, in agreement with the finding in \cite{sc12}.
Consistent results were also found when we made the measurements with data from
each 16-MHz sub-band instead of using the full bandwidth.

\begin{figure}
\centering
\includegraphics[scale=0.36,angle=-90]{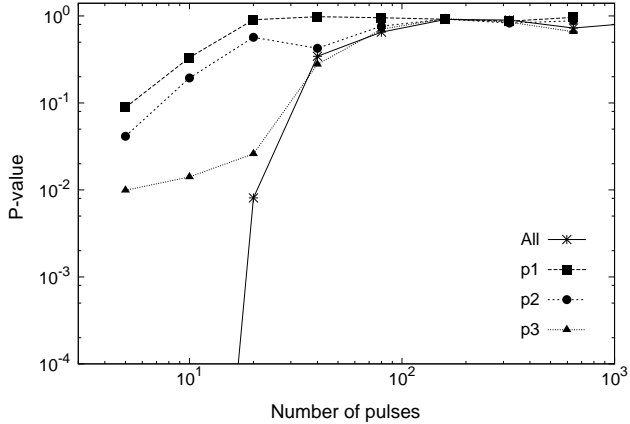}
\caption{Measured $p$-values from standard K-S test of weighted
timing residuals based on different numbers of averaged pulses,
obtained from all pulses as well as sub-sets of pulses based on
their peak flux densities. Details of the three sub-sets can be
found in Table.~\ref{tab:rms}. \label{fig:pv}}
\end{figure}

The phase jitter of integrated profiles is usually assumed to follow
a Gaussian distribution. However, if single pulses are not normally
distributed in rotational phase, the resulting jitter noise is
likely to deviate from Gaussian noise, especially when the number of
averaged pulses is not sufficiently large. For PSR~J1713+0747, the
asymmetric occurrence rate of single pulses in rotational phase is clear
from Fig.~\ref{fig:phavf}, which was also shown in
\cite{sc12}. To further investigate the nature of the jitter
noise, we produced timing residuals based on different
numbers of averaged pulses, weighted them by the sum of radiometer
and jitter noise, and calculated their $p$-values\footnote{Such a
$p$-value close to unity shows that we cannot detect significant
deviation from the normal distribution of data.} from a standard K-S test
under the hypothesis of a normal distribution (Fig.~\ref{fig:pv}). We can see that the
phase jitter starts to appear as Gaussian noise only after integrating
$\sim10^2$ pulses. Integrating a sub-set of pulses selected based on a chosen
range of peak flux densities gives a relatively quick conversion, as
expected. This places a lower limit on the number of pulses (i.e., $\sim10^2$) to be
integrated in timing analysis to avoid significant
non-Gaussianity. Note that a similar limit was already derived for PSR~J0437$-$4715, based on
simulations of the regularity of profile shapes \citep{lvk+11}.

Single pulses do not necessarily have the same variability as a
function of rotational phase \citep[e.g.,][]{cstt96,lkl+15}, meaning
that their contribution to the resulting jitter noise can be
different. In \cite{ovb+14}, single pulses from PSR~J0437$-$4715
were grouped based on their peak flux density and averaged into
integrated profiles, which did not result in a significant
improvement of the timing precision. Following this, we divided the
pulses into groups based on their relative peak flux densities,
relative total flux densities and widths, separately. Subsequently,
we produced timing residuals containing the same number of TOAs from
each group\footnote{The TOAs were estimated with the standard
template matching algorithm \citep{tay92}. To time each sub-set of
pulses, we formed the template by integrating the whole sub-set and
performed wavelet-smoothing to the integrated profile, to avoid
self-correlation by the noise. The residuals were calculated with
the ephemeris from \cite{dcl+16}, without any fit for the timing
parameters.}. The results are summarized in Table~\ref{tab:rms}. It
can be seen that all sub-sets contribute approximately the same
jitter noise per pulse, though pulses of high flux density and
narrow width tend to show slightly lower values. The achieved timing
rms values are also significantly higher than when using all pulses.
Thus, there is no apparent benefit in grouping the single pulses for
the purpose of precision timing.

\begin{table}
\centering \caption{Timing residuals achieved with different sub-sets
  of pulses. The pulses are grouped with respect to their relative
  peak flux density ($p$, peak flux density divided by its running
  mean), relative flux density ($f$, total flux density divided by its
  running mean) and pulse width ($w$, defined as $W_{\rm 50}$).
  The rms values are calculated based on the same number of TOAs.} \label{tab:rms}
\begin{tabular}[c]{cccccc}
\hline
&Range &Percent &RMS ($\mu$s) &$\chi^2_{\rm re}$ &$\sigma_{\rm j,0}$ ($\mu$s)\\
\hline \hline
$p1$ &$(0, 1.17)$       &35.4    &1.01  &2.75 &22.5\\
$p2$ &$(1.17, 1.60)$    &31.3    &1.02  &8.20 &25.1\\
$p3$ &$(1.60, 3.00)$    &29.3    &0.816 &18.9 &20.2\\
$p4$ &$(3.00, +\infty)$ &3.93    &1.97  &69.4 &18.3\\
\hline
$f1$ &$(0, 0.77)$   &31.7    &1.04  &2.87 &22.1\\
$f2$ &$(0.77, 1.05)$   &26.6    &1.11  &9.98 &25.4\\
$f3$ &$(1.05, 1.39)$   &25.7    &0.862 &10.5 &19.5\\
$f4$ &$(1.39, +\infty)$   &16.0    &0.935 &18.5 &17.1\\
\hline
$w1$ &$(0, 30)\,\mu$s   &33.9    &1.06 &3.70 &21.1\\
$w2$ &$(30, 50)\,\mu$s   &35.5    &0.833  &11.3 &21.2\\
$w3$ &$(50, +\infty)\,\mu$s   &30.6    &0.867 &17.8 &21.2\\
\hline
All  &$(0, +\infty)$   &100     &0.590 &12.9 &26.5\\
\hline
\end{tabular}
\end{table}

In principle, the achievable TOA precision should be independent of the
choice of time resolution ($\delta\tau$) in forming the integrated
profile, as far as its features are fully resolved
\citep[e.g.,][]{dr83}. Here we used the single pulse data to form
integrated profiles of 1-s integrations and variable time resolution,
and calculated their corresponding TOA uncertainties. On average, the
TOA uncertainty drops no more than 5\% when $\delta\tau$ decreases
from 8.9\,$\mu$s to 140\,ns. This suggests that the narrow features in the
single pulses are mostly averaged out, and a time resolution beyond
10\,$\mu$s is sufficient in shaping the profiles when the
exposure time is longer than one second.

\section{Conclusions and discussions} \label{sec:conclu}
We have studied the properties of single pulses from PSR~J1713+0747
using 15 minutes of data (corresponding to $\sim197,000$ pulses)
collected by LEAP when the pulsar was extremely bright. The pulse
energy distribution is shown to be consistent with a log-normal
distribution. The pulse widths are typically around 0.04\,ms for the
bright pulses. In addition, we have confirmed the detection of
periodic intensity modulation by \cite{es03a}, and in addition
revealed its association with systematic drifting sub-pulses. From
the 2DFS, $P_2$ and $P_3$ of the two modes were measured to be
($15^{+2}_{-6}$\,deg, $6.9\pm0.1$\,$P$) and ($23^{+2}_{-14}$\,deg,
$2.9\pm0.1$\,$P$), respectively. The occurrence of the two modes is
found to overlap at the phase of the main peak in the integrated
profile, providing the first evidence for superposed modes of
drifting sub-pulse in MSPs. The mode at 0.14\,cpp was also apparent
at the trailing edge of the integrated profile, where the other was
not detected. We did not find any evidence of periodic pulse
micro-structure with a time resolution of up to 140\,ns. With full
polarisation data, we have shown that the bright pulses are
significantly linearly polarised. On average, the fractional
polarisation of the pulses increases with increasing pulse peak flux
density. Using single pulse polarisation, we have presented a
longitude-resolved probability density map of P.A., and shown the
existence of two orthogonal modes of polarisation that are clearly
distinct in pulse phase. Finally, we found that jitter noise induced
by pulse variability can be described as Gaussian only when at least
100 pulses have been integrated. Statistically, pulses of different
properties contribute equally to the resulting jitter noise, though
pulses with high peak flux density have slightly less effect. By
timing sub-sets of pulses with respect to their relative peak flux
density, relative total flux density and pulse width, we did not
find any significant improvement on the overall timing precision. Changing
the time resolution used on 1-s integrations from 8.9\,$\mu$s to 140\,ns
does not result in significant difference in their TOA uncertainties, meaning
that most features on those integrated profiles are resolved with time resolution
beyond 10\,$\mu$s.

Micro-structure in pulsed radio emission has mostly been searched
for in canonical pulsars, and most of the searches that have
sufficient time resolution and sensitivity have achieved successful
detection \citep[e.g.][]{lkwj98,mar15}. On the other hand, besides
PSR~J1713+0747, micro-structure has also been searched for in
another two MSPs, PSR~J0437$-$4715 and PSR~B1937+21, with no
detection \citep{jak+98,jap01}. Using observations of canonical
pulsars, \cite{kjv02} established a relation between micro-structure
width and pulsar rotational period. A simple extrapolation from that
relation leads to a micro-structure width of $0.5-10$\,$\mu$s for
MSPs, well above the time resolutions of the data used for the
searches in MSPs. Thus, the non-detections may imply that either
micro-structure is less common in MSPs compared with the situation
in canonical pulsars, or the relation between micro-structure width and rotational
period is different in MSPs. Nevertheless, this needs to be further verified
by more sample studies.

The discovery of drifting sub-pulses in PSR~J1713+0747 increases the
number of MSP ``drifters'' to three, together with PSR~J1012+5307
and J1518+4904 \citep{es03a}. As high-sensitivity searches for
drifting sub-pulses have been performed in no more than ten MSPs,
drifting sub-pulses do not seem to be an unusual phenomenon among
this part of the pulsar population. It is interesting to note that
all three exhibit a quasi-periodic drifting fashion, and are
classified as ``diffuse'' drifter by the definition in \cite{wes06}.
Given that nearly half of the drifters in canonical pulsars show a
precise drifting period (referred to as ``coherent'' drifters),
quasi-periodic drifting may be comparatively more common in MSPs.
Again, robust statistics of the ratio requires future work including
an expanded number of samples. Searching for drifting sub-pulses in more
MSPs will also show if the phenomenon is correlated with any pulsar properties, such
as the characteristic age which has been discovered among the population of non-recycled pulsars \citep{wes06,wse07}.

Despite the vast increase in sensitivity delivered by LEAP, a
thorough understanding of single pulse properties, in particular
those from the lower end of the energy distribution, is still
limited by S/N. The situation is likely to be significantly improved
when coherent addition of the pulse signal is achieved with more
telescopes, as also suggested in \cite{dlc+14}. Eventually, the next
generation of radio telescopes, e.g., the Square Kilometre Array and
the Five hundred meter Aperture Spherical Telescope, will provide
the best opportunity ever to study single pulse emission from MSPs.

The behaviour of pulse variability is seen to be a source-dependent
phenomenon among pulsars including MSPs. Despite the lack of
improvement in the timing precision of PSR~J1713+0747, single pulse
data could still be used to mitigate jitter noise in some other
pulsars, especially when pulses with different properties exhibit
different distributions in phase. In this case, a weighting scheme
concerning their contribution to jitter noise may be introduced when
integrating the pulses, so as to optimally use the signal of the
pulses which show less variability in phase. An approach in a
similar framework has already been discussed in \cite{ick+15}.
Nevertheless, implementation of such methods is yet to be thoroughly
investigated.

\section*{Acknowledgements}
We would like to thank S.~Os{\l}owski, A.~Noutsos and J.~M.~Cordes for
valuable discussions, I.~Cognard for assistance in observations at Nan\c{c}ay,
and A.~Jessner for carefully reading the manuscript and providing valuable insights. We also
thank the anonymous referee for providing constructive suggestions to improve the article.
K.~Liu acknowledges the financial support by the European Research Council for the ERC Synergy
Grant BlackHoleCam under contract no. 610058. This work is supported by the ERC Advanced Grant
``LEAP", Grant Agreement Number 227947 (PI M.~Kramer). The Effelsberg
100-m telescope is operated by the Max-Planck-Institut f\"{u}r
Radioastronomie. The Westerbork Synthesis Radio Telescope is operated
by the Netherlands Foundation for Radio Astronomy, ASTRON, with
support from NWO. The Nan\c{c}ay Radio Observatory is operated by the
Paris Observatory, associated with the French Centre National de la
Recherche Scientifique. CGB and SS acknowledges support from the European
Research Council under the European Union's Seventh Framework
Programme (FP/2007-2013) / ERC Grant Agreement nr. 337062 (DRAGNET; PI
Jason Hessels). RS acknowledges the financial support from the European
Research Council under the European Union's Seventh Framework Programme
(FP/2007-2013) / ERC Grant Agreement n. 617199. KJL gratefully acknowledges
support from National Basic Research Program of China, 973 Program, 2015CB857101
and NSFC U15311243. This research is a result of
the common effect to directly detect gravitational waves using pulsar timing,
known as the European Pulsar Timing Array (EPTA).

\bibliographystyle{mnras}
\bibliography{journals_apj,psrrefs,modrefs,crossrefs}

\begin{thebibliography}{}

\bibitem[\protect\citeauthoryear{Alpar et~al.}{Alpar et~al.}{1982}]{acrs82}
Alpar M.~A., Cheng A.~F., Ruderman M.~A.,  Shaham J., 1982, Nature, 300, 728

\bibitem[\protect\citeauthoryear{{Antoniadis} et~al.}{{Antoniadis}
  et~al.}{2013}]{afw+13}
{Antoniadis} J. et~al., 2013, Science, 340, 448

\bibitem[\protect\citeauthoryear{{Arzoumanian} et~al.}{{Arzoumanian}
  et~al.}{2015}]{abb+15}
{Arzoumanian} Z. et~al., 2015, ApJ, 813, 65

\bibitem[\protect\citeauthoryear{{Arzoumanian} et~al.}{{Arzoumanian}
  et~al.}{2016}]{abb+16}
{Arzoumanian} Z. et~al., 2016, ApJ, 821, 13

\bibitem[\protect\citeauthoryear{Bartel \& Sieber}{Bartel \&
  Sieber}{1978}]{bs78}
Bartel N.,  Sieber W., 1978, A\&A, 70, 260

\bibitem[\protect\citeauthoryear{{Bassa} et~al.}{{Bassa} et~al.}{2016}]{bjk+16}
{Bassa} C.~G. et~al., 2016, MNRAS, 456, 2196

\bibitem[\protect\citeauthoryear{{Bilous}}{{Bilous}}{2012}]{bil12}
{Bilous} A.~V., 2012, Ph.D. thesis, University of Virginia

\bibitem[\protect\citeauthoryear{{Bilous} et~al.}{{Bilous}
  et~al.}{2015}]{bpd+15}
{Bilous} A.~V., {Pennucci} T.~T., {Demorest} P.,  {Ransom} S.~M., 2015, ApJ,
  803, 83

\bibitem[\protect\citeauthoryear{{Caballero} et~al.}{{Caballero}
  et~al.}{2016}]{cll+16}
{Caballero} R.~N. et~al., 2016, MNRAS, 457, 4421

\bibitem[\protect\citeauthoryear{{Cairns}, {Johnston}, \& {Das}}{{Cairns}
  et~al.}{2004}]{cjd04}
{Cairns} I.~H., {Johnston} S.,  {Das} P., 2004, MNRAS, 353, 270

\bibitem[\protect\citeauthoryear{Cheng \& Ruderman}{Cheng \&
  Ruderman}{1977}]{cr77}
Cheng A.~F.,  Ruderman M., 1977, ApJ, 212, 800

\bibitem[\protect\citeauthoryear{Cognard et~al.}{Cognard et~al.}{1996}]{cstt96}
Cognard I., Shrauner J.~A., Taylor J.~H.,  Thorsett S.~E., 1996, ApJ, 457, L81

\bibitem[\protect\citeauthoryear{{Cordes}}{{Cordes}}{1976}]{cor76a}
{Cordes} J.~M., 1976, ApJ, 210, 780

\bibitem[\protect\citeauthoryear{Cordes}{Cordes}{1979}]{cor79}
Cordes J.~M., 1979, Space Sci. Rev., 24, 567

\bibitem[\protect\citeauthoryear{Cordes \& Downs}{Cordes \& Downs}{1985}]{cd85}
Cordes J.~M.,  Downs G.~S., 1985, ApJS, 59, 343

\bibitem[\protect\citeauthoryear{Craft, Comella, \& Drake}{Craft
  et~al.}{1968}]{ccd68}
Craft H.~D., Comella J.~M.,  Drake F., 1968, Nature, 218, 1122

\bibitem[\protect\citeauthoryear{{Demorest} et~al.}{{Demorest}
  et~al.}{2010}]{dpr+10}
{Demorest} P.~B., {Pennucci} T., {Ransom} S.~M., {Roberts} M.~S.~E.,  {Hessels}
  J.~W.~T., 2010, Nature, 467, 1081

\bibitem[\protect\citeauthoryear{{Desvignes} et~al.}{{Desvignes}
  et~al.}{2011}]{dbc+11}
{Desvignes} G., {Barott} W.~C., {Cognard} I., {Lespagnol} P.,  {Theureau} G.,
  2011, in American Institute of Physics Conference Series, Vol. 1357, {Burgay}
  M., {D'Amico} N., {Esposito} P., {Pellizzoni} A.,  {Possenti} A., ed,
  American Institute of Physics Conference Series, p. 349

\bibitem[\protect\citeauthoryear{{Desvignes} et~al.}{{Desvignes}
  et~al.}{2016}]{dcl+16}
{Desvignes} G. et~al., 2016, MNRAS, 458, 3341

\bibitem[\protect\citeauthoryear{{Dolch} et~al.}{{Dolch} et~al.}{2014}]{dlc+14}
{Dolch} T. et~al., 2014, ApJ, 794, 21

\bibitem[\protect\citeauthoryear{Downs \& Reichley}{Downs \&
  Reichley}{1983}]{dr83}
Downs G.~S.,  Reichley P.~E., 1983, ApJS, 53, 169

\bibitem[\protect\citeauthoryear{Drake \& Craft}{Drake \& Craft}{1968}]{dc68}
Drake F.~D.,  Craft H.~D., 1968, Nature, 220, 231

\bibitem[\protect\citeauthoryear{{Edwards} \& {Stappers}}{{Edwards} \&
  {Stappers}}{2003}]{es03a}
{Edwards} R.~T.,  {Stappers} B.~W., 2003, A\&A, 407, 273

\bibitem[\protect\citeauthoryear{{Edwards} \& {Stappers}}{{Edwards} \&
  {Stappers}}{2004}]{es04}
{Edwards} R.~T.,  {Stappers} B.~W., 2004, A\&A, 421, 681

\bibitem[\protect\citeauthoryear{{Freire} et~al.}{{Freire}
  et~al.}{2012}]{fwe+12}
{Freire} P.~C.~C. et~al., 2012, MNRAS, 423, 3328

\bibitem[\protect\citeauthoryear{{Gil}}{{Gil}}{1985}]{gil85}
{Gil} J., 1985, ApJS, 110, 293

\bibitem[\protect\citeauthoryear{{Gil}, {Melikidze}, \& {Geppert}}{{Gil}
  et~al.}{2003}]{gmg03}
{Gil} J., {Melikidze} G.~I.,  {Geppert} U., 2003, A\&A, 407, 315

\bibitem[\protect\citeauthoryear{Gil \& Lyne}{Gil \& Lyne}{1995}]{gl95}
Gil J.~A.,  Lyne A.~G., 1995, MNRAS, 276, L55

\bibitem[\protect\citeauthoryear{{Hankins}}{{Hankins}}{1972}]{han72}
{Hankins} T.~H., 1972, ApJ, 177, L11

\bibitem[\protect\citeauthoryear{{Imgrund} et~al.}{{Imgrund}
  et~al.}{2015}]{ick+15}
{Imgrund} M., {Champion} D.~J., {Kramer} M.,  {Lesch} H., 2015, MNRAS, 449,
  4162

\bibitem[\protect\citeauthoryear{{Janssen} et~al.}{{Janssen}
  et~al.}{2015}]{jhm+15}
{Janssen} G. et~al., 2015, Advancing Astrophysics with the Square Kilometre
  Array (AASKA14), 37

\bibitem[\protect\citeauthoryear{Jenet et~al.}{Jenet et~al.}{1998}]{jak+98}
Jenet F., Anderson S., Kaspi V., Prince T.,  Unwin S., 1998, ApJ, 498, 365

\bibitem[\protect\citeauthoryear{{Jenet}, {Anderson}, \& {Prince}}{{Jenet}
  et~al.}{2001}]{jap01}
{Jenet} F.~A., {Anderson} S.~B.,  {Prince} T.~A., 2001, ApJ, 546, 394

\bibitem[\protect\citeauthoryear{{Karastergiou} et~al.}{{Karastergiou}
  et~al.}{2002}]{kkj+02}
{Karastergiou} A., {Kramer} M., {Johnston} S., {Lyne} A.~G., {Bhat} N.~D.~R.,
  {Gupta} Y., 2002, A\&A, 391, 247

\bibitem[\protect\citeauthoryear{{Karuppusamy}, {Stappers}, \& {van
  Straten}}{{Karuppusamy} et~al.}{2008}]{ksv08}
{Karuppusamy} R., {Stappers} B.,  {van Straten} W., 2008, PASP, 120, 191

\bibitem[\protect\citeauthoryear{{Knight}}{{Knight}}{2007}]{kni07}
{Knight} H.~S., 2007, MNRAS, 378, 723

\bibitem[\protect\citeauthoryear{{Knight} et~al.}{{Knight}
  et~al.}{2006}]{kbm+06b}
{Knight} H.~S., {Bailes} M., {Manchester} R.~N., {Ord} S.~M.,  {Jacoby} B.~A.,
  2006, ApJ, 640, 941

\bibitem[\protect\citeauthoryear{{Kramer}, {Johnston}, \& {van
  Straten}}{{Kramer} et~al.}{2002}]{kjv02}
{Kramer} M., {Johnston} S.,  {van Straten} W., 2002, MNRAS, 334, 523

\bibitem[\protect\citeauthoryear{Kramer et~al.}{Kramer et~al.}{2006}]{ksm+06}
Kramer M. et~al., 2006, Science, 314, 97

\bibitem[\protect\citeauthoryear{Lange et~al.}{Lange et~al.}{1998}]{lkwj98}
Lange C., Kramer M., Wielebinski R.,  Jessner A., 1998, A\&A, 332, 111

\bibitem[\protect\citeauthoryear{{Lazarus} et~al.}{{Lazarus}
  et~al.}{2016}]{lkg+16}
{Lazarus} P., {Karuppusamy} R., {Graikou} E., {Caballero} R.~N., {Champion}
  D.~J., {Lee} K.~J., {Verbiest} J.~P.~W.,  {Kramer} M., 2016, MNRAS, 458, 868

\bibitem[\protect\citeauthoryear{{Lentati} \& {Shannon}}{{Lentati} \&
  {Shannon}}{2015}]{ls15}
{Lentati} L.,  {Shannon} R.~M., 2015, MNRAS, 454, 1058

\bibitem[\protect\citeauthoryear{{Lentati} et~al.}{{Lentati}
  et~al.}{2015}]{ltm+15}
{Lentati} L. et~al., 2015, MNRAS, 453, 2576

\bibitem[\protect\citeauthoryear{{Liu} et~al.}{{Liu} et~al.}{2015}]{lkl+15}
{Liu} K. et~al., 2015, MNRAS, 449, 1158

\bibitem[\protect\citeauthoryear{{Liu} et~al.}{{Liu} et~al.}{2012}]{lkl+11}
{Liu} K., {Keane} E.~F., {Lee} K.~J., {Kramer} M., {Cordes} J.~M.,  {Purver}
  M.~B., 2012, MNRAS, 420, 361

\bibitem[\protect\citeauthoryear{{Liu} et~al.}{{Liu} et~al.}{2011}]{lvk+11}
{Liu} K., {Verbiest} J.~P.~W., {Kramer} M., {Stappers} B.~W., {van Straten} W.,
   {Cordes} J.~M., 2011, MNRAS, 417, 2916

\bibitem[\protect\citeauthoryear{{Lyne} et~al.}{{Lyne} et~al.}{2010}]{lhk+10}
{Lyne} A., {Hobbs} G., {Kramer} M., {Stairs} I.,  {Stappers} B., 2010, Science,
  329, 408

\bibitem[\protect\citeauthoryear{{Lyne} \& {Graham-Smith}}{{Lyne} \&
  {Graham-Smith}}{2006}]{lg06}
{Lyne} A.~G.,  {Graham-Smith} F., 2006, {Pulsar Astronomy}

\bibitem[\protect\citeauthoryear{McKinnon \& Stinebring}{McKinnon \&
  Stinebring}{1998}]{ms98}
McKinnon M.,  Stinebring D., 1998, ApJ, 502, 883

\bibitem[\protect\citeauthoryear{{Mitra}, {Arjunwadkar}, \& {Rankin}}{{Mitra}
  et~al.}{2015}]{mar15}
{Mitra} D., {Arjunwadkar} M.,  {Rankin} J.~M., 2015, ApJ, 806, 236

\bibitem[\protect\citeauthoryear{{Mitra}, {Gil}, \& {Melikidze}}{{Mitra}
  et~al.}{2009}]{mgm09}
{Mitra} D., {Gil} J.,  {Melikidze} G.~I., 2009, ApJ, 696, L141

\bibitem[\protect\citeauthoryear{{Ord} et~al.}{{Ord} et~al.}{2004}]{ovhb04}
{Ord} S.~M., {van Straten} W., {Hotan} A.~W.,  {Bailes} M., 2004, MNRAS, 352,
  804

\bibitem[\protect\citeauthoryear{{Os{\l}owski} et~al.}{{Os{\l}owski}
  et~al.}{2014}]{ovb+14}
{Os{\l}owski} S., {van Straten} W., {Bailes} M., {Jameson} A.,  {Hobbs} G.,
  2014, MNRAS, 441, 3148

\bibitem[\protect\citeauthoryear{{Os{\l}owski} et~al.}{{Os{\l}owski}
  et~al.}{2011}]{ovh+11}
{Os{\l}owski} S., {van Straten} W., {Hobbs} G.~B., {Bailes} M.,  {Demorest} P.,
  2011, MNRAS, 418, 1258

\bibitem[\protect\citeauthoryear{{{\"O}zel} et~al.}{{{\"O}zel}
  et~al.}{2010}]{opr+10}
{{\"O}zel} F., {Psaltis} D., {Ransom} S., {Demorest} P.,  {Alford} M., 2010,
  ApJ, 724, L199

\bibitem[\protect\citeauthoryear{{Qiao} et~al.}{{Qiao} et~al.}{2004}]{qlz+04}
{Qiao} G.~J., {Lee} K.~J., {Zhang} B., {Xu} R.~X.,  {Wang} H.~G., 2004, ApJ,
  616, L127

\bibitem[\protect\citeauthoryear{Rankin}{Rankin}{1986}]{ran86}
Rankin J.~M., 1986, ApJ, 301, 901

\bibitem[\protect\citeauthoryear{{Reardon} et~al.}{{Reardon}
  et~al.}{2016}]{rhc+16}
{Reardon} D.~J. et~al., 2016, MNRAS, 455, 1751

\bibitem[\protect\citeauthoryear{Ruderman \& Sutherland}{Ruderman \&
  Sutherland}{1975}]{rs75}
Ruderman M.~A.,  Sutherland P.~G., 1975, ApJ, 196, 51

\bibitem[\protect\citeauthoryear{{Serylak}, {Stappers}, \&
  {Weltevrede}}{{Serylak} et~al.}{2009}]{ssw09}
{Serylak} M., {Stappers} B.~W.,  {Weltevrede} P., 2009, A\&A, 506, 865

\bibitem[\protect\citeauthoryear{{Shannon} \& {Cordes}}{{Shannon} \&
  {Cordes}}{2012}]{sc12}
{Shannon} R.~M.,  {Cordes} J.~M., 2012, ApJ, 761, 64

\bibitem[\protect\citeauthoryear{{Shannon} et~al.}{{Shannon}
  et~al.}{2014}]{sod+14}
{Shannon} R.~M. et~al., 2014, MNRAS, 443, 1463

\bibitem[\protect\citeauthoryear{{Shannon} et~al.}{{Shannon}
  et~al.}{2015}]{srl+15}
{Shannon} R.~M. et~al., 2015, Science, 349, 1522

\bibitem[\protect\citeauthoryear{{Simmons} \& {Stewart}}{{Simmons} \&
  {Stewart}}{1985}]{ss85}
{Simmons} J.~F.~L.,  {Stewart} B.~G., 1985, A\&A, 142, 100

\bibitem[\protect\citeauthoryear{Stairs, Thorsett, \& Camilo}{Stairs
  et~al.}{1999}]{stc99}
Stairs I.~H., Thorsett S.~E.,  Camilo F., 1999, ApJS, 123, 627

\bibitem[\protect\citeauthoryear{Stinebring et~al.}{Stinebring
  et~al.}{1984}]{scr+84}
Stinebring D.~R., Cordes J.~M., Rankin J.~M., Weisberg J.~M.,  Boriakoff V.,
  1984, ApJS, 55, 247

\bibitem[\protect\citeauthoryear{Taylor}{Taylor}{1992}]{tay92}
Taylor J.~H., 1992, Phil. Trans. Roy. Soc. A, 341, 117

\bibitem[\protect\citeauthoryear{{Tiburzi} et~al.}{{Tiburzi}
  et~al.}{2013}]{tjb+13}
{Tiburzi} C. et~al., 2013, MNRAS, 436, 3557

\bibitem[\protect\citeauthoryear{van Straten}{van Straten}{2004}]{van04a}
van Straten W., 2004, ApJ, 152, 129

\bibitem[\protect\citeauthoryear{{van Straten}}{{van Straten}}{2010}]{van10}
{van Straten} W., 2010, ApJ, 719, 985

\bibitem[\protect\citeauthoryear{{van Straten} \& {Bailes}}{{van Straten} \&
  {Bailes}}{2011}]{vb11}
{van Straten} W.,  {Bailes} M., 2011, Proc. Astr. Soc. Aust., 28, 1

\bibitem[\protect\citeauthoryear{{Verbiest} et~al.}{{Verbiest}
  et~al.}{2016}]{vlh+16}
{Verbiest} J.~P.~W. et~al., 2016, MNRAS, 458, 1267

\bibitem[\protect\citeauthoryear{Wardle \& Kronberg}{Wardle \&
  Kronberg}{1974}]{wk74}
Wardle J.,  Kronberg P., 1974, ApJ, 194, 249

\bibitem[\protect\citeauthoryear{{Weisberg}, {Nice}, \& {Taylor}}{{Weisberg}
  et~al.}{2010}]{wnt10}
{Weisberg} J.~M., {Nice} D.~J.,  {Taylor} J.~H., 2010, ApJ, 722, 1030

\bibitem[\protect\citeauthoryear{{Weltevrede}}{{Weltevrede}}{2016}]{wel16}
{Weltevrede} P., 2016, A\&A, 590, A109

\bibitem[\protect\citeauthoryear{{Weltevrede}, {Edwards}, \&
  {Stappers}}{{Weltevrede} et~al.}{2006}]{wes06}
{Weltevrede} P., {Edwards} R.~T.,  {Stappers} B.~W., 2006, A\&A, 445, 243

\bibitem[\protect\citeauthoryear{{Weltevrede}, {Stappers}, \&
  {Edwards}}{{Weltevrede} et~al.}{2007}]{wse07}
{Weltevrede} P., {Stappers} B.~W.,  {Edwards} R.~T., 2007, A\&A, 469, 607

\bibitem[\protect\citeauthoryear{{Weltevrede} et~al.}{{Weltevrede}
  et~al.}{2006}]{wws+06}
{Weltevrede} P., {Wright} G.~A.~E., {Stappers} B.~W.,  {Rankin} J.~M., 2006,
  A\&A, 458, 269

\bibitem[\protect\citeauthoryear{Xilouris et~al.}{Xilouris
  et~al.}{1994}]{xkjw94}
Xilouris K.~M., Kramer M., Jessner A.,  Wielebinski R., 1994, A\&A, 288, L17

\bibitem[\protect\citeauthoryear{{Yan} et~al.}{{Yan} et~al.}{2011}]{ymv+11}
{Yan} W.~M. et~al., 2011, MNRAS, 414, 2087

\bibitem[\protect\citeauthoryear{{Zhu} et~al.}{{Zhu} et~al.}{2015}]{zsd+15}
{Zhu} W.~W. et~al., 2015, ApJ, 809, 41

\bibitem[\protect\citeauthoryear{{Zhuravlev} et~al.}{{Zhuravlev}
  et~al.}{2013}]{zps+13}
{Zhuravlev} V.~I., {Popov} M.~V., {Soglasnov} V.~A., {Kondrat'ev} V.~I.,
  {Kovalev} Y.~Y., {Bartel} N.,  {Ghigo} F., 2013, MNRAS, 430, 2815

\end{thebibliography}
\end{document}